%% file: main.tex
\documentclass[acmtog]{acmart}

\renewcommand\footnotetextcopyrightpermission[1]{}

\AtBeginDocument{%
  \fancypagestyle{firstpagestyle}{%
    \fancyhf{}%
  }%
  \fancypagestyle{standardpagestyle}{%
    \fancyhf{}%
  }%
  \pagestyle{standardpagestyle}}

\usepackage{amsmath}
\usepackage{booktabs}
\usepackage{placeins}
\usepackage{dblfloatfix}
\usepackage{xcolor}
\usepackage{pdfpages}


\usepackage[normalem]{ulem}
\begin{document}
\setcopyright{none}

\input{utils}

\title[A Unified Conditional Flow for Motion Generation, Editing, and Intra-Structural Retargeting]{A Unified Conditional Flow for Motion Generation, Editing, and Intra-Structural Retargeting}

\author{Junlin Li}
\affiliation{%
  \institution{ByteDance}
  \city{San Jose}
  \country{USA}}
\email{1306043330a@gmail.com}
\author{Xinhao Song}
\affiliation{%
  \institution{ByteDance}
  \city{San Jose}
  \country{USA}}
\email{xinhao.song13@gmail.com}
\author{Siqi Wang}
\affiliation{%
  \institution{ByteDance}
  \city{San Jose}
  \country{USA}}
\email{siqi9494@gmail.com}
\author{Haibin Huang}
\affiliation{%
  \institution{ByteDance}
  \city{San Jose}
  \country{USA}}
\email{jackiehuanghaibin@gmail.com}
\author{Yili Zhao}
\affiliation{%
  \institution{ByteDance}
  \city{San Jose}
  \country{USA}}
\email{yilizhao.cs@gmail.com}

\begin{abstract}
Text-driven motion editing and intra-structural retargeting, where skeletons share topology but may differ in bone lengths and rest pose, are traditionally handled by fragmented pipelines with incompatible inputs and representations: editing relies on specialized generative steering, while retargeting is deferred to geometric post-processing. We present a unified conditional-flow framework that casts generation, semantic editing, and intra-structural retargeting as \emph{condition-modulated transport} within one text- and skeleton-conditioned rectified-flow model. Under this formulation, editing changes the semantic condition while preserving skeletal structure, whereas retargeting changes the skeletal condition while preserving motion semantics. This makes FlowEdit-style transport a unified inference rule for motion manipulation rather than a task-specific editor. To instantiate this for articulated 3D motion, we develop a text- and skeleton-conditioned rectified-flow transformer. The model uses per-joint tokenization and explicit joint self-attention to capture spatial kinematic dependencies. We further inject text conditions at both joint and frame levels, while residual multi-condition classifier-free guidance balances text adherence and skeletal conformity.

Experiments on SnapMoGen and a multi-character Mixamo subset show that one trained model supports text-to-motion generation, zero-shot editing, and zero-shot intra-structural retargeting without task-specific fine-tuning. This unified framework replaces separate pipelines with a single conditional motion transport model while keeping the same-topology retargeting scope explicit.
\end{abstract}


\begin{teaserfigure}
  \centering
  \includegraphics[width=\linewidth]{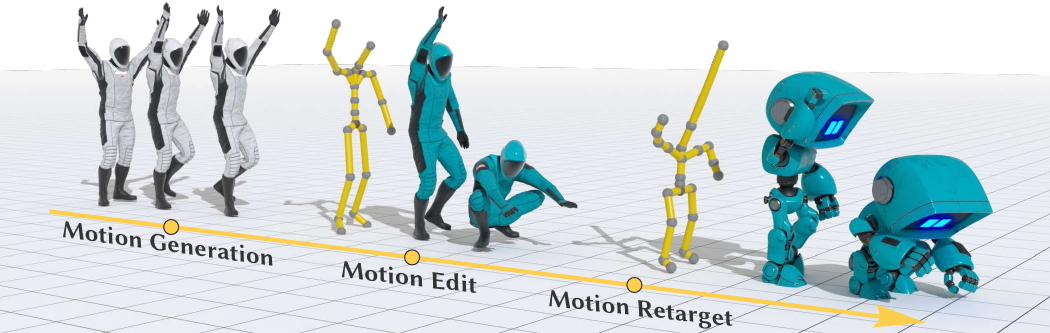}
  \caption{\textbf{One rectified-flow model unifies motion generation, editing, and intra-structural retargeting.}
Conditioned on text and skeleton, it enables (left) generation,
(middle) zero-shot editing by changing only the text condition, and (right)
zero-shot retargeting by changing only the skeleton condition.}
  \label{fig:teaser}
\end{teaserfigure}

\maketitle

\input{body}

\begin{acks}
We thank Chongyang Ma and Yizhi Wang for helpful discussions and feedback. We thank Bingxuan Li for assistance with exploratory experiments. We are also grateful to Kai Hu for his strong support of this project.
\end{acks}

\bibliographystyle{ACM-Reference-Format}
\bibliography{ref}

\clearpage
\input{images}

\clearpage
\includepdf[pages=-]{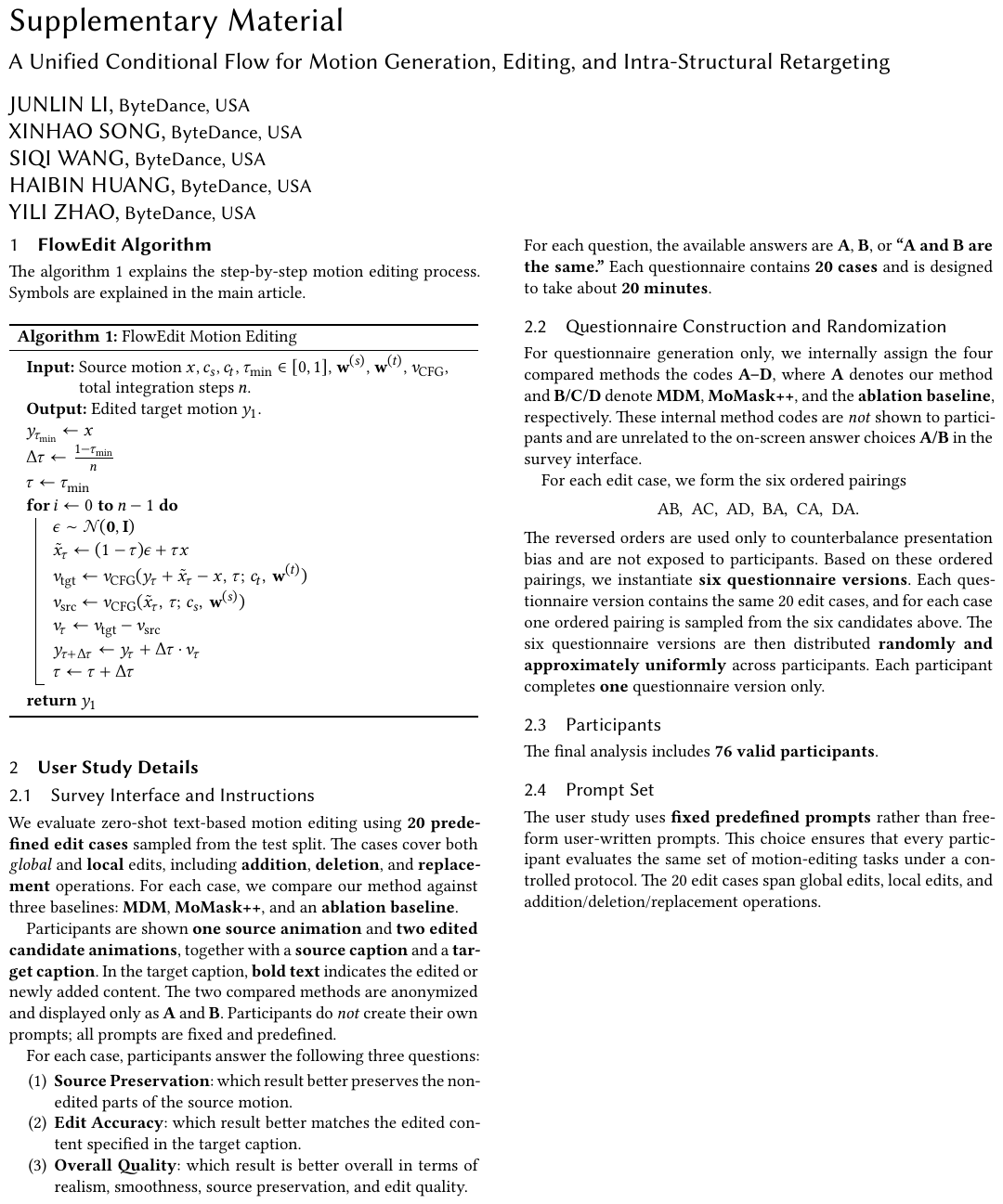}

\end{document}

%% file: utils.tex
\definecolor{darkBlue}{RGB}{0, 94, 184}
\definecolor{iglBlue}{RGB}{144,210,236}
\definecolor{iglGreen}{RGB}{153,203,67}
\definecolor{coralRed}{RGB}{250,114,104}
\definecolor{gray}{RGB}{180,180,180}
\definecolor{orange}{RGB}{255,165,0}
\definecolor{TechnionBlue}{RGB}{8,33,78}
\definecolor{Purple}{RGB}{137, 99, 198}
\definecolor{lightgray}{gray}{0.65}
\definecolor{babyblue}{RGB}{203, 218, 235}
\definecolor{red}{RGB}{255, 0, 0}
\newcommand{\modified}[1]{#1}

%% file: body.tex
\input{intro}
\input{related_work}
\input{method}
\input{exp}
\input{conclusion}

%% file: intro.tex
\section{INTRODUCTION}
Character animation relies heavily on reusing and modifying existing motion data. Traditionally cross-character reuse is handled by motion retargeting; here we study intra-structural retargeting, where characters share skeleton topology but differ in proportions. Recently, text-to-motion models enable \textit{prompt-driven motion editing}, allowing high-level semantic changes in natural language~\citep{tevet2023mdm,chen2023mld,zhang2022motiondiffuse,guo2024momask,guo2025snapmogen}. However, the two remain separate: editing typically relies on scarce edit supervision~\citep{athanasiou2024motionfix,li2025simmotionedit} or heuristic steering~\citep{hong2025salad}, while retargeting is still treated as a post-process relying on inverse kinematics or learned mappings~\citep{gleicher1998retargetting,aberman2020san,zhang2023r2et}.

This fragmentation stems from the fact that editing and retargeting are usually formulated with incompatible inputs, objectives, and representations. First, the inputs differ: editing conditions on a \emph{source motion} and a \emph{text prompt}, whereas retargeting conditions on a \emph{source motion} and a \emph{target skeleton}. Second, the objectives diverge: text-driven editing seeks to preserve the source text-motion alignment while altering its semantics, whereas retargeting seeks to preserve the semantic intent while adapting the motion to a new skeletal morphology. Third, and most critically, the underlying representations are mismatched. Most text-to-motion generators operate in a canonical output space (e.g., a fixed kinematic tree with standardized bone lengths) to simplify training~\citep{tevet2023mdm,chen2023mld}. Consequently, motion adaptation is deferred to downstream tools rather than handled by the generative model itself. This separation complicates deployment—requiring multiple models and inconsistent representations—and hinders the composition of operations, such as simultaneous editing and retargeting, within a single interactive system.

We propose a unified conditional-flow perspective that resolves this fragmentation. Both tasks are recast as instances of condition-modulated transport within a single rectified-flow model conditioned jointly on text and skeleton. The key insight is a structural symmetry between the two: editing changes the text condition while holding the skeleton fixed; retargeting changes the skeleton condition while holding the text fixed. Building on FlowEdit ~\citep{kulikov2025flowedit}, which constructs a direct source-to-target transport ODE without explicit inversion, this symmetry yields a unified inference rule, the same update equation, the same trained model, applied to both tasks by simply swapping which condition changes.

Bringing this principle to articulated 3D motion is \emph{nontrivial}. Unlike pixels, motion involves joint rotations under kinematic constraints, strong inter-joint dependencies, temporal coherence, and the competing need to simultaneously produce benchmark compatible generation features and geometry-faithful retargeting features-challenges that do not arise in image or video domains.

We address these with a single rectified-flow model trained on clips paired with text and skeleton information, using the target T-pose as a compact encoding of bone lengths and rest pose. The backbone is a DiT-style joint-token transformer~\citep{peebles2022dit} that represents each motion as a concatenated generation/retargeting feature pair, applies explicit joint self-attention within each frame to model bone-length-dependent kinematic dependencies, and injects text at both joint and frame resolution via dual cross-attention. Shape and semantic adherence during sampling are controlled independently through residual multi-condition classifier-free guidance, with separate skeleton and text weights.

At inference time, the same model supports all three tasks through one update rule: generation samples from a joint text--skeleton condition, text editing changes only the prompt while keeping the skeleton fixed, and intra-structural retargeting changes only the skeleton while preserving source motion semantics. No task-specific modules or fine-tuning are required. The same trained model and inference rule support text-to-motion generation, zero-shot text editing, and zero-shot intra-structural retargeting. We evaluate the framework on SnapMoGen~\citep{guo2025snapmogen} and a multi-character Mixamo subset across all those three tasks. On SnapMoGen, our unified multi-skeleton version ($0.787/0.893/0.929$ R-Precision, 17.148 FID) nearly matches our single-skeleton variant ($0.807/0.906/0.939$, 16.493), showing minimal cost from skeleton conditioning. On Mixamo, we achieve the best retargeting results: $4.67\times10^{-3}$ joint-position error, 0.961 contact accuracy, and 1.23 cm/s foot-skating. User studies confirm preference for our edits, demonstrating that one conditional-flow model unifies generation, editing, and retargeting.

To summarize, our contribution is two-fold:
\begin{itemize}
    \item We introduce a unified system design where one conditional-flow model handles generation, zero-shot editing, and intra-structural retargeting under a single conditional-flow paradigm.

    \item To our knowledge, this is the first framework to bring FlowEdit style transport to 3D character motion, adapting it to articulated skeletons with kinematic constraints, inter-joint dependencies, and temporal coherence.
\end{itemize}

%% file: related_work.tex
\section{RELATED WORK}
\label{sec:related_work}

\subsection{Text-to-Human Motion Generation}
\label{sec:related_work_generation}
Text-driven motion generation has advanced with larger motion--language datasets \cite{guo2022humanml3d,guo2025snapmogen,fan2025gotzero} and stronger generative models. Token-based methods compress motion into VQ-style codes \cite{oord2018vqvae} and model them autoregressively or with masked prediction~\cite{zhang2023t2mgpt,guo2024momask,guo2025snapmogen}, while diffusion and flow models avoid some quantization artifacts~\cite{tevet2023mdm,zhang2022motiondiffuse,huang2024stablemofusion,meng2024mardm,hong2025salad,gao2025motionflux,he2026molingo,wen2025hymotion,li2026motionhiflow}. Recent work also studies per-joint streaming latents, action planning, compound actions, coordinate autoregression, and recognition-guided generation~\cite{li2026llamo,ling2026prism,nazarenus2026actionplan,jiang2026motionadapter,ding2026cdamd,kuang2026coamd}. Yet most methods still sample in a fixed canonical skeleton space and defer motion adaptation to later retargeting. Skeletal control has instead been explored through skeleton-conditioned generation/retargeting~\cite{cao2025gdream}, topology- or skeleton-adaptive generation~\cite{gat2025anytop,liu2025omniskel}, kinematic constraints~\cite{rempe2026kimodo}, runtime smart primitives~\cite{wang2026motionbricks}, and identity/body shape~\cite{jia2026iam}. Like prior motion transformers, we also use spatial joint reasoning~\cite{aksan2021spatiotemporal}; our difference is that we merge the joint tokens into frame tokens to leverage the feed-forward blocks, and use the target skeleton as a first-class condition, so it can be changed during sampling and editing.

\subsection{Motion Editing}
\label{sec:related_work_editing}
Motion editing modifies an input sequence while preserving unedited content, timing, and motion identity. Supervised text-driven methods learn from source--instruction--target triplets~\cite{athanasiou2024motionfix,li2025simmotionedit}, but such data are scarce, motivating zero-shot editors based on inpainting, conditioning control, or latent/code refinement~\cite{tevet2023mdm,chen2025motionclr,hong2025salad,huang2024como,jeong2025pgr2m}. Our method is most closely related to FlowEdit~\cite{kulikov2025flowedit} and to attention- or inversion-based image/video editing~\cite{hertz2022prompttoprompt,qi2023fatezero}. Recent motion systems broaden motion generation/editing toward unified or instruction-following interfaces, motion-condition modeling, part-level composition/editing, multi-human interactions, and unified motion synthesis/understanding~\cite{jiang2026motionmaster,guo2025motionlab,bu2025omnimogen,cong2026umo,yang2025partmotionedit,li2026frankenmotion,yang2026interedit,li2024unimotion}. We adapt inversion-free conditional transport to articulated motion and extend conditioning from text alone to text plus skeleton, so semantic and skeleton editing share one transport equation rather than two task-specific inference pipelines.

\subsection{Motion Retargeting}
\label{sec:related_work_retargeting}
Motion retargeting transfers motion across morphologically different characters while preserving semantic intent and physical plausibility. Learning-based methods often cast retargeting as unpaired translation between skeleton domains with cycle or adversarial objectives~\cite{villegas2018nkn,aberman2020san,lim2019pmnet}, while skeleton-agnostic embeddings broaden skeleton support~\cite{lee2023same}. Beyond skeleton-only transfer, mesh-aware methods emphasize geometry for contact, interpenetration, and body-part relationships through voxel distance fields~\cite{zhang2023r2et}, semantically consistent sensors~\cite{ye2024meshret}, and surface-derived constraints~\cite{zhang2025smrnet}. Recent systems~\cite{chen2025motion2motion,kim2025moreflow,zhang2026adamorph,gong2026mocapanything,mueller2026reactor,saito2026soma,nvidia2026somaretargeter} broaden the scope to cross-topology transfer, unsupervised flow-matching retargeting, arbitrary rigs, physics-aware robot retargeting, and heterogeneous body-model/robot ecosystems. These broader systems often rely on explicit correspondences, IK, or optimization. Our setting is complementary: we study skeleton level, intra-structural retargeting inside the generator, where changing the skeleton condition invokes the same conditional-flow transport used for text editing. We compare against retargeting baselines to measure target-skeleton conformity and source-motion preservation.

%% file: method.tex
\section{METHOD}
\label{sec:method}

We propose treating motion retargeting as a conditional \emph{generative} problem rather than a geometric optimization task. Under this view, the target skeleton (defined by bone lengths and rest pose) serves as a conditioning signal analogous to a ``style'' or ``domain'' code, while the motion semantics serve as the ``content.'' This parallels the logic of text-driven editing: just as changing a text prompt modifies the semantic trajectory of a motion, changing a skeleton condition modifies its morphological trajectory. By leveraging the FlowEdit framework~\citep{kulikov2025flowedit} on top of a unified generative backbone, we enable both text-driven semantic edits and structure-driven retargeting within a single model, without requiring task-specific modules or training.

\begin{figure*}[!t]
  \centering
  \includegraphics[width=0.98\textwidth]{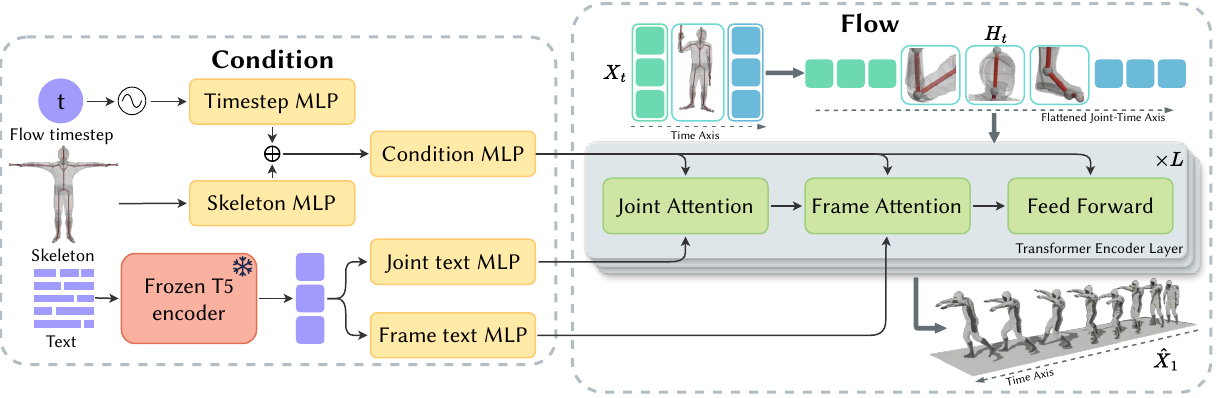}
  \Description{Model architecture diagram for the dual-conditioned transformer with text and skeleton conditioning.}
  \caption{Model Architecture. Input frame tokens are reshaped into per-joint tokens for processing. Time and skeleton conditions are injected via AdaLN, while text embeddings are integrated through cross-attention.}
  \label{fig:pipeline}
\end{figure*}

To achieve this vision, our approach relies on three integrated components: a dual-conditioned transformer backbone, a conditional rectified flow training strategy, and a unified FlowEdit inference scheme, as summarized in Fig.~\ref{fig:pipeline} and Fig.~\ref{fig:flowedit_tasks}. In the subsequent sections, we first formalize the problem definition and data representation, followed by detailed descriptions of the model architecture and the unified inference procedure.

\subsection{Problem Formulation}
\label{subsec:problem_formulation}

\paragraph{Problem Definition}
We represent a human motion clip as a sequence of feature vectors $\mathbf{x} \in \mathbb{R}^{T \times D}$, where $T$ is the sequence length and $D$ is the feature dimension. Each motion $\mathbf{x}$ is associated with two conditioning signals:
(i) a \textbf{semantic condition} $\mathbf{c}_{\text{txt}}$, derived from a natural language prompt, and
(ii) a \textbf{skeletal condition} $\mathbf{c}_{\text{skel}}$, representing the character's skeletal morphology (e.g., bone lengths and rest pose).

Our objective is to learn a \emph{single} conditional distribution \mbox{\(p(\mathbf{x}\mid \mathbf{c}_{\text{txt}}, \mathbf{c}_{\text{skel}})\)} realized by a shared conditional-flow model for generation, text editing, and intra-structural retargeting.

Specifically, the model must support three operations:
\begin{enumerate}
    \item \textbf{Text-to-Motion Generation:}\newline
    Sampling \(\mathbf{x} \sim \mbox{\(p(\mathbf{x} \mid \mathbf{c}_{\text{txt}}, \mathbf{c}_{\text{skel}})\)}\) for any target skeleton $\mathbf{c}_{\text{skel}}$ within the supported topology.
    \item \textbf{Zero-Shot Text-Based Editing:} Given a source motion $\mathbf{x}_{\text{src}}$ and its conditions, generating a modified motion $\mathbf{x}_{\text{tgt}}$ that adheres to a new prompt $\mathbf{c}_{\text{txt}}^{\text{tgt}}$ while preserving the non-edited part of the $\mathbf{c}_{\text{txt}}^{\text{src}}$ and the original motion.
    \item \textbf{Zero-Shot Intra-Structural Retargeting:} Given a source motion $\mathbf{x}_{\text{src}}$ generated with skeleton $\mathbf{c}_{\text{skel}}^{\text{src}}$, producing a new motion $\mathbf{x}_{\text{tgt}}$ that preserves the source semantics but conforms strictly to a new target skeleton $\mathbf{c}_{\text{skel}}^{\text{tgt}}$ (where the skeletons share a graph topology but differ in bone lengths).
\end{enumerate}

 \textit{Data Representation.}
To satisfy both benchmark compatibility and geometry-faithful retargeting, we represent each motion state as a single concatenated vector: 
 \begin{equation}
   \mathbf{x}_t = \big[\mathbf{x}^{\text{gen}}_t \,;\, \mathbf{x}^{\text{ret}}_t\big],
   \qquad
   \mathbf{x}^{\text{gen}}_t\in\mathbb{R}^{D_{\text{gen}}},\;
   \mathbf{x}^{\text{ret}}_t\in\mathbb{R}^{D_{\text{ret}}}.
   \label{eq:concat_features}
 \end{equation}
 Note that both slices are predicted jointly by the same rectified-flow field and jointly manipulated by the same FlowEdit trajectory. The concatenation is therefore not a task split; it is a dual-view representation of the same motion sample.

 \textit{Generative Features ($\mathbf{x}^{\text{gen}}$).}
The first slice follows the standard SnapMoGen layout ($D_{\text{gen}} = 8 + 12J$), consisting of root velocities (yaw/planar) and height, local joint rotations, root-relative joint positions and velocities, and foot contacts. We retain this slice because standard text-to-motion benchmarks 
are defined on this benchmark-compatible interface. However, it is a poor retargeting interface: root orientation is represented through angular velocities rather than absolute angles, and rotations are temporally smoothed. When motion is transferred to a different skeleton, this representation discards the absolute spatial frame and accumulates reconstruction error, which we observe as root drift.

\textit{Retargeting Features ($\mathbf{x}^{\text{ret}}$).}
To obtain a geometry-faithful retargeting interface, we append AnyTop-style~\citep{gat2025anytop} features ($D_{\text{ret}} = 12J$) for each joint $j \in \{1,\dots,J\}$:
\begin{equation}
  \mathbf{x}^{\text{ret}}_{t,j}
  =
  \big[\mathbf{p}_{t,j}\,;\,\mathbf{r}^{6\text{D}}_{t,j}\,;\,\mathbf{v}_{t,j}\big]
  \in \mathbb{R}^{12}.
  \label{eq:anytop_features}
\end{equation}
 Here, $\mathbf{p}_{t,j}$ denotes canonicalized root-relative joint positions, $\mathbf{v}_{t,j}$ denotes velocities and $\mathbf{r}^{6\text{D}}_{t,j}$ represents the 6D rotation \cite{zhou20196drotation}. Unlike using the root angular velocities to reconstruct the root transformations in $\mathbf{x}^{\text{gen}}$, our root rotation is directly obtained from $\mathbf{r}^{6\text{D}}_{t,j}$. To accurately predict $\mathbf{p}_{t,j}$ conditioned on a skeleton vector $\mathbf{c}_{\text{skel}}$, the model must learn the underlying kinematic constraints (i.e., bone lengths). This \textit{implicit skeleton modeling} is the foundation of our retargeting capability: it allows the FlowEdit inference process to continuously "warp" the skeleton geometry from source to target. 

The two slices serve different downstream interfaces. Standard text-to-motion benchmark metrics are computed from $\mathbf{x}^{\text{gen}}$. Retargeting geometry is reconstructed from $\mathbf{x}^{\text{ret}}$, and unless otherwise stated, final retargeted animations are rendered by applying the predicted $\mathbf{x}^{\text{ret}}$ rotations to the fixed target skeleton via forward kinematics (FK). Foot-contact cues remain available from $\mathbf{x}^{\text{gen}}$ for contact analysis and post-processing.
 
 \textit{Condition Inputs.}
 We parameterize the semantic condition $\mathbf{c}_{\text{txt}}$ using text tokens $\boldsymbol{P} = \{w_i\}_{i=1}^L$ from a frozen T5 encoder.
 The structural condition $\mathbf{c}_{\text{skel}}$ is parameterized as a vector $\boldsymbol{S} \in \mathbb{R}^{D_{\text{ret}}}$ following the layout of Eq.~\ref{eq:anytop_features}. In practice, we set the position channels of $\boldsymbol{S}$ to the target character's T-pose offsets (defining bone lengths and rest pose) and set the remaining rotation/velocity channels to identity/zero.

\subsection{Model Architecture}
We adopt a DiT-style transformer backbone that is jointly conditioned on text and skeleton.
Let $\mathbf{X}\in\mathbb{R}^{T\times D}$ denote an input state such as $\mathbf{x}_\tau$, and let $J$ be the joint count.
Unlike prior motion transformers that treat each frame feature as a single token, we encourage body-part-level generation by operating also on \emph{per-joint tokens}.
Concretely, instead of projecting $\mathbf{X}_t$ to an arbitrary hidden vector, we project it to $J$ joint chunks and reshape:
\begin{equation}
  \mathbf{H}_{t} = \mathrm{reshape}\left(\mathbf{W}_{\text{in}}\mathbf{X}_{t} + \mathbf{b}_{\text{in}}\right)\in\mathbb{R}^{J\times d},
  \qquad
  d = \frac{D_h}{J},
  \label{eq:joint_tokens}
\end{equation}
where $D_h$ is the model total hidden size and $\mathbf{H}_{t,j}$ is the token for joint $j$ at frame $t$.
The input projection learns how to split the whole feature dimensions into joint parts.

\textit{Joint attention.}
Standard transformer backbones capture temporal structure via attention between frame tokens, while spatial inter-joint structure is typically handled \emph{indirectly} by the per-token feed-forward network that mixes all joint parts inside the hidden frame vector.
Without a strong skeletal inductive bias, this mixing must learn inter-joint relations through a generic MLP, which is parameter-heavy and often slow to learn.

We keep temporal attention and add dense joint self-attention over per-frame joint tokens. Related spatial-temporal joint attention has been used for motion prediction \cite{aksan2021spatiotemporal}; in contrast, our joint attention is placed before frame attention and coupled with joint-/frame-level text injection inside a text- and skeleton-conditioned rectified-flow model.

Within each layer, we apply joint self-attention independently at each time step:
\begin{equation}
  \mathbf{H}_{t,\cdot} \leftarrow \mathbf{H}_{t,\cdot} +
  g_{\text{jnt}}(\mathbf{c}) \cdot
  \mathrm{Attn}_{\text{jnt}}\!\left(\mathrm{AdaLN}(\mathbf{H}_{t,\cdot}, \mathbf{c})\right),
  \label{eq:self_joint_attn}
\end{equation}
where $\mathbf{c}$ is the conditioning vector defined below, $\mathrm{Attn}_{\text{jnt}}$ attends over joints $j=1\dots J$ using joint rotary embeddings, and $g_{\text{jnt}}(\mathbf{c})$ is the AdaLN-Zero residual gate.
This directly models intra-frame joint dependencies, which is critical when generating on characters with different bone lengths.

\textit{Frame attention and feed-forward.}
After updating joint tokens, we form frame tokens back by concatenating all joints per frame,
$\mathbf{F}_t = [\mathbf{H}_{t,1};\dots;\mathbf{H}_{t,J}] \in \mathbb{R}^{D_h}$, and run temporal attention over the motion sequence with frame rotary embeddings.
Finally, a SwiGLU feed-forward network updates frame tokens and the outputs are reshaped back to joints.

\textit{Skeleton and flow-time condition injection.}
We encode flow time $\tau$ with a sinusoidal embedding and map it to $\mathbf{c}_\tau\in\mathbb{R}^{D_h}$ with an MLP.
We encode the target T-pose vector with a skeleton MLP, yielding $\mathbf{c}_s\in\mathbb{R}^{D_h}$.
The final conditioning vector is
\begin{equation}
  \mathbf{c} = \phi\!\left(\mathbf{c}_\tau + \mathbf{c}_s\right)\in\mathbb{R}^{D_h},
  \label{eq:cond_vector}
\end{equation}
where $\phi$ is a small condition-merging MLP.
We inject $\mathbf{c}$ into every transformer block using AdaLN-Zero modulation with learned shift, scale, and a residual gate.
This conditioning controls both joint-level and frame-level updates; for example, it drives the gate $g_{\text{jnt}}(\mathbf{c})$ in Eq.~\ref{eq:self_joint_attn}.

\textit{Text injection at joint and frame resolutions.}
We encode text prompt tokens $\{w_i\}_{i=1}^L$ with a frozen T5 encoder, producing token embeddings $\mathbf{e}_i\in\mathbb{R}^{d_{\text{text}}}$.
We then form:
(i) \emph{frame-level} text tokens $\mathbf{E}^{\text{frm}}\in\mathbb{R}^{L\times D_h}$ and
(ii) \emph{joint-level} text tokens $\mathbf{E}^{\text{jnt}}\in\mathbb{R}^{L\times d}$
via two separate linear projections.
After joint attention, joint tokens query the text via cross-attention using $\mathbf{E}^{\text{jnt}}$ as keys/values, while $\mathbf{E}^{\text{frm}}$ participates in the subsequent frame-text cross attention after the frame attention.
This design injects language at both the spatial resolution over joints and the temporal resolution over frames.

\textit{Output head.}
The final joint tokens are normalized and reshaped back to $D_h$, then projected to the motion feature dimension:
\begin{equation}
  \hat{\mathbf{x}}_1 = \mathbf{W}_{\text{out}}\,
  \mathrm{reshape}\!\left(\mathrm{Norm}(\mathbf{H})\right) + \mathbf{b}_{\text{out}}.
  \label{eq:output_head}
\end{equation}
We initialize $\mathbf{W}_{\text{out}},\mathbf{b}_{\text{out}}$ to zero to stabilize early training.

\subsection{Rectified Flow Matching}
We model motions with a rectified flow (RF) \cite{liu2022rectifiedflow} defined on the continuous-time path
$\mathbf{x}_\tau\in\mathbb{R}^{T\times D}$, $\tau\in[0,1]$.
Let $\mathbf{x}_1$ be a data sample and $\mathbf{x}_0\sim\mathcal{N}(0, \boldsymbol{I})$ be base noise.
To match FlowEdit's assumption that the noisy state is a \emph{linear interpolant}, we use rectified flow matching with the linear path
\begin{equation}
  \mathbf{x}_\tau = (1-\tau)\mathbf{x}_0 + \tau\mathbf{x}_1,
  \label{eq:rf_interpolate}
\end{equation}
whose ground-truth velocity is
\begin{equation}
  \mathbf{u}^\star = \frac{d\mathbf{x}_\tau}{d\tau} = \mathbf{x}_1 - \mathbf{x}_0.
  \label{eq:rf_velocity_gt}
\end{equation}

\textit{Predicting the clean target.}
Instead of predicting velocity, we predict the clean target sample $\hat{\mathbf{x}}_1=f_\theta(\mathbf{x}_\tau,\tau; \boldsymbol{P},\boldsymbol{S})$ following the previous work \cite{tevet2023mdm, cuba2025flowmotion}, 
which empirically yields smoother outputs.
To support inference and editing, we then convert the target prediction to a velocity field via
\begin{equation}
  \mathbf{v}_\theta(\mathbf{x}_\tau,\tau; \boldsymbol{P},\boldsymbol{S})
  = \frac{\hat{\mathbf{x}}_1 - \mathbf{x}_\tau}{1-\tau},
  \label{eq:rf_target_to_velocity}
\end{equation}
where $\tau$ is clamped to at most $0.999$ for numerical stability.

\textit{Flow matching loss.}
Training minimizes a mean-squared error on the predicted target directly.
Because $\mathbf{x}_\tau$ concatenates $\mathbf{x}_\tau^{\text{gen}}$ and $\mathbf{x}_\tau^{\text{ret}}$ blocks, we enforce the objective \emph{per feature block}:
\begin{equation}
  \begin{aligned}
    \mathcal{L}_{\text{simple}}
    =\;& \lambda_{\text{gen}}\mathbb{E}_{\mathbf{x}_1,\mathbf{x}_0,\tau}
    \left[\left\lVert \mathbf{f}^{\text{gen}}_\theta(\mathbf{x}_\tau,\tau; \boldsymbol{P},\boldsymbol{S}) - \mathbf{x}_1^{\text{gen}} \right\rVert^2\right] \\
    &+ \lambda_{\text{ret}}\mathbb{E}_{\mathbf{x}_1,\mathbf{x}_0,\tau}
    \left[\left\lVert \mathbf{f}^{\text{ret}}_\theta(\mathbf{x}_\tau,\tau; \boldsymbol{P},\boldsymbol{S}) - \mathbf{x}_1^{\text{ret}} \right\rVert^2\right],
  \end{aligned}
  \label{eq:rf_loss}
\end{equation}
where $\mathbf{f}^{\text{gen}}_\theta$ and $\mathbf{f}^{\text{ret}}_\theta$ are the corresponding slices of the concatenated prediction.
This prevents the higher-dimensional block from implicitly dominating the loss signal.
We set $\lambda_{\text{gen}}=\lambda_{\text{ret}}=1$ unless otherwise stated.

\textit{Multi-condition classifier-free guidance.}
We use multi-condition CFG over the branches exposed by
condition dropout. Since the network predicts the clean target, each branch
output is first converted to a velocity using
Eq.~\ref{eq:rf_target_to_velocity}. We evaluate
\[
\mathbf{v}_0 =
\mathbf{v}_\theta(\mathbf{x}_\tau,\tau;\emptyset,\mathbf{0}),\quad
\mathbf{v}_s =
\mathbf{v}_\theta(\mathbf{x}_\tau,\tau;\emptyset,\boldsymbol{S}),\quad
\mathbf{v}_{ts} =
\mathbf{v}_\theta(\mathbf{x}_\tau,\tau;\boldsymbol{P},\boldsymbol{S}),
\]
corresponding to unconditional, skeleton-only, and text+skeleton
conditioning. The guided velocity is
\begin{equation}
\mathbf{v}_{\mathrm{CFG}}
=
\mathbf{v}_0
+ w_{\mathrm{skel}}(\mathbf{v}_s-\mathbf{v}_0)
+ w_{\mathrm{text}}(\mathbf{v}_{ts}-\mathbf{v}_s).
\label{eq:v_multi_cfg}
\end{equation}

Here \(w_{\mathrm{skel}}\) controls \emph{shape} adherence by moving from the unconditional field to the skeleton-conditioned field, while \(w_{\mathrm{text}}\) controls \emph{semantic} adherence as a residual correction on top of the skeleton-conditioned field. 
We therefore omit a text-only branch \(\mathbf{v}_\theta(\mathbf{x}_\tau,\tau;\boldsymbol{P},\mathbf{0})\), because text guidance is defined relative to the skeleton-conditioned field. During training, condition dropout uses \(p_{\mathrm{both}}=p_{\mathrm{text}}=0.1\): dropping both trains \(\mathbf{v}_0\), dropping text trains \(\mathbf{v}_s\), and dropped text/skeleton are encoded as padding tokens/zero vectors.

\begin{table*}[t]
\caption{Text-to-motion generation on SnapMoGen. Baselines follow SnapMoGen (mean$\pm$95\% CI); bold/underline denote best/second best. ``Ours w/o Mixamo Split'' trains on the same SnapMoGen-only data as the baselines, with our model setup.}
\label{tab:gen}
\centering
\small
{\setlength{\tabcolsep}{6pt}
\begin{tabular*}{\textwidth}{@{\extracolsep{\fill}}p{0.30\textwidth}ccc ccc@{}}
\toprule
& \multicolumn{3}{c}{R Precision$\uparrow$} & & & \\
\cmidrule(lr){2-4}
Methods & Top 1 & Top 2 & Top 3 & FID$\downarrow$ & CLIP Score$\uparrow$ & MModality$\uparrow$ \\
\midrule
MDM & 0.503$\pm$0.002 & 0.653$\pm$0.002 & 0.727$\pm$0.002 & 57.783$\pm$0.092 & 0.481$\pm$0.001 & \textbf{13.412$\pm$0.231} \\
T2M-GPT & 0.618$\pm$0.002 & 0.773$\pm$0.002 & 0.812$\pm$0.002 & 32.629$\pm$0.087 & 0.573$\pm$0.001 & 9.172$\pm$0.181 \\
StableMoFusion & 0.679$\pm$0.002 & 0.823$\pm$0.002 & 0.888$\pm$0.002 & 27.801$\pm$0.063 & 0.605$\pm$0.001 & 9.064$\pm$0.138 \\
MARDM & 0.659$\pm$0.002 & 0.812$\pm$0.002 & 0.860$\pm$0.002 & 26.878$\pm$0.131 & 0.602$\pm$0.001 & 9.812$\pm$0.287 \\
MoMask & 0.777$\pm$0.002 & 0.888$\pm$0.002 & 0.927$\pm$0.002 & 17.404$\pm$0.051 & 0.664$\pm$0.001 & 8.183$\pm$0.184 \\
MoMask++ & \uline{0.802$\pm$0.001} & \uline{0.905$\pm$0.002} & \uline{0.938$\pm$0.001} & \textbf{15.060$\pm$0.065} & \uline{0.685$\pm$0.001} & 7.259$\pm$0.180 \\
\midrule
\textbf{Ours} & 0.787$\pm$0.003 & 0.893$\pm$0.002 & 0.929$\pm$0.002 & 17.148$\pm$0.059 & 0.676$\pm$0.001 & \uline{10.51$\pm$0.201} \\
\textbf{Ours, w/o Mixamo Split} & \textbf{0.807$\pm$0.002} & \textbf{0.906$\pm$0.002} & \textbf{0.939$\pm$0.001} & \uline{16.493$\pm$0.082} & \textbf{0.688$\pm$0.001} & 9.580$\pm$0.176 \\
\bottomrule
\end{tabular*}
}
\end{table*}

\begin{figure}[t]
  \centering
  \includegraphics[width=\columnwidth]{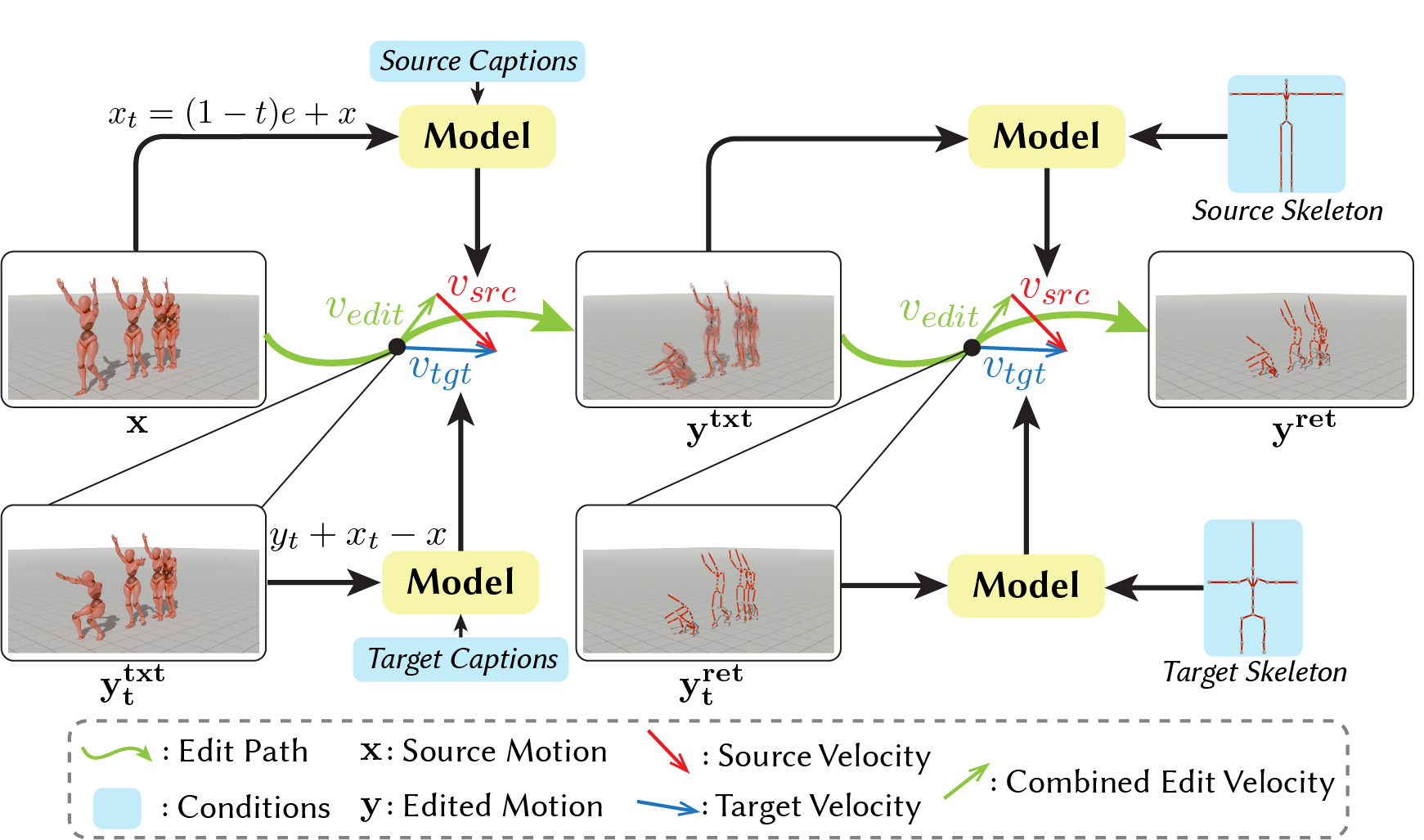}
  \Description{Text editing and retargeting use the same inversion-free conditional-flow update.}
  \caption{Unified FlowEdit inference scheme. Left: Text editing changes the text condition; Right: Intra-structural retargeting changes the skeleton condition. Both use the same trained model and inversion-free update.}
  \label{fig:flowedit_tasks}
\end{figure}

\subsection{FlowEdit for Zero-Shot Editing and Retargeting}
FlowEdit \cite{kulikov2025flowedit} is an inversion-free conditional transport method, originally used for text-based image/video editing. 
It transports a source sample toward a target condition by integrating the difference between source- and target-conditioned velocity fields along a shared noised source path. 
These velocity fields are obtained from an existing well-trained flow model that works on the current edited states. 
Unlike SDEdit and ODEdit, which edit by adding noise to the input and re-running the generative process, it avoids the explicit inversion or round-trip through pure noise. 
This construction of a direct, shorter vector trajectory in latent space between the source and target distribution empirically lowers the transport cost and preserves more source details.
\textit{Intuitively}, zero-shot editing/retargeting becomes possible because the trained model already defines velocities for different text and skeleton conditions. The directions where the source and target share the same semantics cancel out, so the source details are retained. The other parts diverge, and the edit happens. A step-by-step FlowEdit algorithm is provided in the supplementary material.

The original FlowEdit formulation parameterizes time as $t\in[0,1]$ with $t=0$ representing data and $t=1$ representing noise.
In this paper we instead use $\tau\in[0,1]$ with $\tau=0$ representing noise and $\tau=1$ representing data, consistent with our rectified-flow interpolation in Eq.~\ref{eq:rf_interpolate}. Let the source motion be \(x\), with source condition
\(c_s=(\boldsymbol{P}_s,\boldsymbol{S}_s)\) and target condition
\(c_t=(\boldsymbol{P}_t,\boldsymbol{S}_t)\). We sample a shared noise
\(\boldsymbol{\epsilon}\sim\mathcal{N}(\mathbf{0},\mathbf{I})\) and define
\begin{equation}
  \tilde{x}_\tau = (1-\tau)\boldsymbol{\epsilon}+\tau x.
  \label{eq:flowedit_noisy_source}
\end{equation}

\textit{Update rule.}
Starting from \(y_{\tau_{\min}}=x\), we integrate
\begin{equation}
  \frac{dy_\tau}{d\tau}
  =
  v_{\mathrm{CFG}}(y_\tau+\tilde{x}_\tau-x,\tau;c_t,\mathbf{w}^{(t)})
  -
  v_{\mathrm{CFG}}(\tilde{x}_\tau,\tau;c_s,\mathbf{w}^{(s)}),
  \label{eq:flowedit_ode}
\end{equation}
where \(\tau\in[\tau_{\min},1]\), \(y_\tau\) is the updated motion, \(c_s\) and \(c_t\) are the source
and target conditions, and
\(\mathbf{w}^{(a)}=(w_{\mathrm{skel}}^{(a)},w_{\mathrm{text}}^{(a)})\),
\(a\in\{s,t\}\), are the source/target CFG weights in
Eq.~\ref{eq:v_multi_cfg}. The two velocity evaluations share
the same \(\tilde{x}_\tau\), and thus the same noise. This is shown in Fig.\ref{fig:flowedit_tasks}.

\textit{Edit strength.}
We initialize at \(\tau_{\min}\geq 0\) so the first target evaluation is
anchored to the same partially noised source state:
\(y_{\tau_{\min}}+\tilde{x}_{\tau_{\min}}-x=\tilde{x}_{\tau_{\min}}\).
Smaller \(\tau_{\min}\) injects more noise and yields stronger edits,
while larger \(\tau_{\min}\) better preserves the source motion.
The source and target weights, \(\mathbf{w}^{(s)}\) and
\(\mathbf{w}^{(t)}\), further control this trade-off: increasing the
relevant target weight relative to its source counterpart strengthens
the requested semantic or structural change, while reducing the gap
preserves more source detail.

\textit{Task instantiations.}
\textbf{(i) Zero-shot text-based editing:} keep the skeleton fixed and change only text,
$c_s=(\boldsymbol{P},\boldsymbol{S})$, $c_t=(\boldsymbol{P}',\boldsymbol{S})$.
\textbf{(ii) Zero-shot intra-structural retargeting:} set text empty and change only skeleton,
$c_s=(\emptyset,\boldsymbol{S})$, $c_t=(\emptyset,\boldsymbol{S}')$.
Both tasks use the same trained model and the same FlowEdit update rule, without extra training or fine-tuning.

%% file: exp.tex
\section{EXPERIMENTS}

\subsection{Dataset}

\textit{SnapMoGen.}
We use SnapMoGen \citep{guo2025snapmogen}, which provides 20,450 motion clips (43.7 hours) and 122,565 descriptions averaging 48 words. Clips are 4--12~s at 30 FPS in a canonical BVH skeleton, so we use its canonical skeleton for all text-to-motion evaluation.

\textit{Mixamo retargeting subset.} Because SnapMoGen has only one skeleton, we build a supplementary multi-character Mixamo subset \citep{adobe2018mixamo} for retargeting. We filter common locomotion/action keywords (e.g., \texttt{run}, \texttt{walk}, \texttt{sit}), keep 30\% of matches to minimize redundancy, and pair each motion with 3 distinct characters to model cross-morphology transfer. We standardize assets to the SnapMoGen topology by normalizing height, pruning incompatible joints (e.g., fingers), and reordering the kinematic chain. We then use Qwen3-VL~\citep{bai2025qwen3vl} to generate 6 captions per motion from the original Mixamo filename as a prompt prior, matching SnapMoGen caption style and length, and merge both sources with an 80/20 train/test split. Representative generated caption examples are provided in the supplementary.

The details of the subset are as follows: 1,546 motions, 108 characters, 4,638 motion-character clips, 9,276 ordered retargeting pairs, 27,828 generated captions, 320-frame training windows, 1,237/309 train/test motions, 86/22 train/test characters, 3,711/927 train/test clips.
The motion frames range from 62 to 1403; the durations are between 2.07 and 46.77s, with a mean of 4.25s.
For the variation of height-normalized characters: leg-length/height: 0.1987-0.5401; arm-span/height: 0.5365-1.1183; shoulder-width/height: 0.0378-0.1692; hip-width/height: 0.0750-0.2023.

\subsection{Implementation details}
\textit{Model Configuration.}
We use a 12-layer transformer with 12 attention heads, feed-forward expansion ratio 3, and dropout 0.1. To align with our per-joint tokenization strategy, we set the total hidden dimension to $D_h = 432$. For the standard topology with $J=24$ joints, this gives $d=18$ channels per joint token ensuring sufficient capacity for the joint self-attention mechanism. Text is encoded with a frozen T5-Large encoder \citep{ni2022t5}.

\textit{Training Protocol.}
We train on fixed 320-frame windows, obtained by cropping or padding clips, using AdamW \citep{loshchilov2019adamw} with learning rate $5\times10^{-5}$, weight decay $5\times10^{-3}$, global batch size 512, and 400 epochs. Training takes about 8 hours on one node with 8 NVIDIA H100 GPUs.

\textit{Motion Post-Processing.}
\label{subsec:motion_post_processing}
Inspired by \citet{saito2026soma}, we stabilize feet using predicted heel/toe contact channels (Sec.~\ref{subsec:problem_formulation}), fix contacted positions, and blend smoothly into and out of those fixed segments to avoid abrupt changes. Then we resolve the lower body with inverse kinematics so the hips and legs stay consistent while reducing foot sliding. Finally we apply a lightweight moving-average filter to root translation and joint rotations to suppress jitter.

\subsection{Text-to-motion generation.}
We evaluate core generation on the SnapMoGen test split using its single canonical skeleton for fair comparison. Following prior work~\citep{tevet2023mdm, guo2025snapmogen}, we report FID, R-Precision (Top 3), CLIP score, and Multimodality in the feature space of a pre-trained TMR model~\citep{petrovich2023tmr}. We sample with fourth-order Runge--Kutta and 100 steps~\citep{hairer1993solvingode1}, using $w_{\text{txt}}=2.0$ and $w_{\text{skel}}=1.0$ for CFG (Eq.~\ref{eq:v_multi_cfg}).

Fig. ~\ref{fig:result-gen} visualizes samples generated from long, complex prompts.
Table~\ref{tab:gen} compares our method against state-of-the-art baselines under two training settings. \emph{Ours w/o Mixamo Split} isolates the architecture from the added multi-character data and achieves the best R-Precision and CLIP score with second-best FID. The full model, trained with the Mixamo split for intra-structural retargeting, is slightly weaker on pure generation but remains competitive, indicating that skeleton-conditioned transport adds retargeting support with little loss in single-skeleton text-to-motion quality.

Crucially, prior methods often trade alignment for diversity. In contrast, our method maintains a robust balance: it delivers state-of-the-art alignment and competitive realism without collapsing diversity. Most importantly, these results are achieved using a \emph{unified} model trained to support varying skeletons within a shared topology. This demonstrates that our dual-conditioning architecture incurs almost no performance penalty on the standard single-skeleton task.

\begin{table}
\caption{Generation ablation on SnapMoGen.}
\label{tab:gen-ablation}
\centering
\small
{\setlength{\tabcolsep}{2pt}
\begin{tabular}{@{}p{0.36\linewidth}ccc@{}}
\toprule
Methods & FID$\downarrow$ & CLIP$\uparrow$ & MM$\uparrow$ \\
\midrule
Ours w/o jnt attn
& 17.465$\pm$0.085 & 0.680$\pm$0.001 & 10.460$\pm$0.216 \\
Ours
& \textbf{17.148$\pm$0.059} & 0.676$\pm$0.001 & \textbf{10.512$\pm$0.201} \\
\bottomrule
\end{tabular}
}
\end{table}

\textit{Ablation Study.} Table~\ref{tab:gen-ablation} shows that removing joint self-attention worsens FID despite only $\sim$0.4M fewer parameters (58M parameters in total), so the gain comes from the \emph{inductive bias} of explicit spatial modeling rather than raw capacity.

\subsection{Motion Editing}
\label{subsec:exp_editing}
We evaluate zero-shot motion editing for semantic change 
while maintaining the structural integrity of the source motion.

\textit{Setup and Inference.}
We test global and local edits---addition, deletion, and replacement---on 20 test motions (randomly sampled from the test split) against MDM \cite{tevet2023mdm}, MoMask++ \cite{guo2025snapmogen}, and a mask-based ablation using MDM's editing protocol. 
The selection of the baselines is based on fairness: our editing is zero-shot with no edit-triplet supervision, task-specific fine-tuning, or manual masks. Some recent methods, including \citet{athanasiou2024motionfix} and \citet{li2025simmotionedit}, require explicit \mbox{supervised} \mbox{source–instruction–target} triplets. MDM and MoMask++ are the closest zero-shot baselines.
We use MDM's unmodified native mask/inpainting editing protocol and follow CoMo \cite{huang2024como}, fusing the source description and requested change into one prompt to regenerate with MoMask++.
We use 100 steps, source/target text guidance $1.5/3.5$, and skeleton guidance $1.0$ on both branches. We control edit strength via starting time $\tau_{\min}\!\approx\!0.1$.

\begin{figure}
\centering
\includegraphics[width=0.46\textwidth]{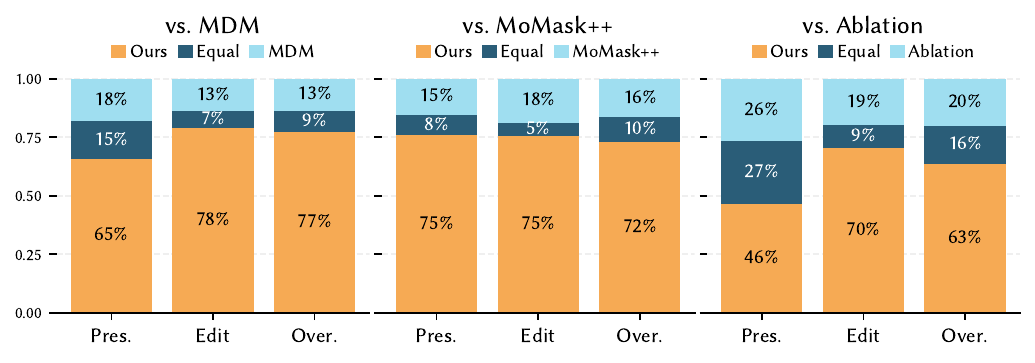}
\caption{\textbf{Perceptual user study.} Pairwise preferences for source preservation, edit accuracy, and overall quality; Ours leads in every criterion.}

\label{fig:user_study}
\end{figure}

\textit{Perceptual Evaluation.}
Fig.~\ref{fig:result-edit} shows qualitative examples. Given the subjective nature of motion editing, we conducted an A/B user study ($N=76$ valid participants).
Participants evaluated 20 motion pairs across 6 survey sets using 3 criteria:
\textbf{Source Preservation}, measuring consistency of unedited parts;
\textbf{Edit Accuracy}, measuring alignment with the target text; and
\textbf{Overall Quality}, measuring realism and smoothness.
Full survey interface and protocol details are provided in supplementary.
Fig.~\ref{fig:user_study} shows that users prefer our method in \textbf{70\%} of comparisons on average. These results demonstrate a robust preference for our flow-based transport over mask-based alternatives. Unlike MDM, our method is fully automatic and requires no manual masking.

\subsection{Motion Retargeting}
We evaluate zero-shot intra-structural retargeting between characters that share topology but significantly differ in proportions.

\textit{Inference Configuration.}
We use a 100-step integrator, sweep the start step from 5 to 40 with stride 5, and set \(w_{\mathrm{skel}}=0.8\), which gives the lowest FK error in our sweep. Unless otherwise stated, text is empty on both FlowEdit branches.

\textit{Baselines \& Metrics.}
We compare against SAN~\citep{aberman2020san} (w/o the adversarial loss), SAME~\citep{lee2023same}, R2ET~\citep{zhang2023r2et}, and a copy baseline. Table~\ref{tab:retargeting_with_post_processing} reports height-normalized global joint-position MSE, foot skating speed, contact accuracy over all heel/toe channels, and strict frame contact accuracy, which requires all four channels to be correct in a frame. 

\textit{Quantitative Results.}
Table~\ref{tab:retargeting_with_post_processing} shows that our full pipeline delivers the strongest overall retargeting performance. The final outputs achieve the best MSE, indicating that the model learns source--target skeletal correspondence and adapts motion faithfully to target proportions rather than merely copying source poses. This is already evident before FK and post-processing: decoding the predicted position channels yields a low MSE of \emph{5.16} (not shown in table), confirming that the network has learned target-conforming geometry in position space rather than relying on FK to impose plausibility afterward. 
The improvement over the no-joint-attention ablation further indicates that joint attention strengthens cross-joint structural reasoning. 
The contact-oriented metrics further show that our full pipeline delivers cleaner and more stable retargeting than competing methods, with lower sliding velocity on predicted static-contact frames and the best per-channel contact accuracy in the final outputs. 
Although \emph{Frame Contact Accuracy} is not the top score, the copy baseline shows that contact agreement alone does not imply successful retargeting; taken together, the metrics indicate the best balance of fidelity and contact stability. Fig.~\ref{fig:result-retarget} further visualizes retargeting outcomes across different methods.

\begin{table}[tbp]
\caption{Intra-structural retargeting on the disjoint Mixamo test split, where both motions and characters are unseen during training. MSE is height-normalized global joint-position error. Foot Skate is mean foot joint speed on predicted static-contact frames. Contact Acc. is over heel/toe channels; Frame Contact Acc. requires all four channels correct in a frame.}
\label{tab:retargeting_with_post_processing}
\centering
\small
\setlength{\tabcolsep}{3pt}
\begin{tabular*}{\columnwidth}{@{\extracolsep{\fill}}lcccc@{}}
\toprule
Method
& \shortstack{MSE\\($\times 10^{3}$)$\downarrow$}
& \shortstack{Foot Skate\\(cm/s)$\downarrow$}
& \shortstack{Contact\\Acc.$\uparrow$}
& \shortstack{Frame Contact\\Acc.$\uparrow$} \\
\midrule
Copy
& 15.89
& 2.803 
& 0.954 
& \textbf{0.898} \\
SAN
& 17.52
& 6.080 
& 0.780
& 0.564 \\
SAME
& 13.47
& 3.965 
& 0.938 
& 0.813 \\
R2ET
& 8.13
& 3.026 
& 0.960
& 0.896 \\
\midrule
Ours w/o jnt attn
& 5.03
& \textbf{1.192}
& 0.958
& 0.859 \\
Ours
& \textbf{4.67}
& 1.230
& \textbf{0.961}
& 0.865 \\
\bottomrule
\end{tabular*}
\end{table}

\textit{Text--Skeleton Interaction in Retargeting.}
We analyze this with a C0/C1/C2 sweep under Eq.~\ref{eq:v_multi_cfg}: C0 uses empty text, C1 the matched source caption, and C2 a length-matched counterfactual caption. Because the same text is used on both FlowEdit branches, this probes text--skeleton/state interaction rather than pure semantic assistance. A small calibrated text term improves FK error near \(w_{\mathrm{text}}=0.25, w_{\mathrm{skel}}=0.8\), but C1 and C2 remain close, indicating limited semantic contribution. We therefore use empty text by default for retargeting and leave stronger semantic--skeletal coupling to future work; details are in the supplementary.

\textit{Generalization to Out-of-distribution Characters.}
Table~\ref{tab:retargeting_with_post_processing} evaluates unseen motions and unseen characters using disjoint Mixamo splits over both motions and characters. We further test external same-topology target assets absent from both SnapMoGen and Mixamo. Despite substantial variations in scale, proportions, and body shape, our method coherently retargets the same source motion to these characters without character-specific fine-tuning (Fig. ~\ref{fig:ood}). Since paired ground truth is unavailable, we report these results qualitatively and provide full sequences in the supplementary video.

We additionally conduct a controlled morphology stress test (Fig. \ref{fig:ood_stress_test}), progressively stretching a same-topology target character from mild to extreme proportions and retargeting the same source motion to each variant. The mildly stretched case (yellow) remains satisfactory, demonstrating generalization to out-of-distribution characters. With larger deformations, artifacts such as ground penetration and altered leg configurations begin to appear, revealing the failure mode under extreme morphology shifts.

\modified{
\begin{figure}
\centering
\includegraphics[width=0.46\textwidth]{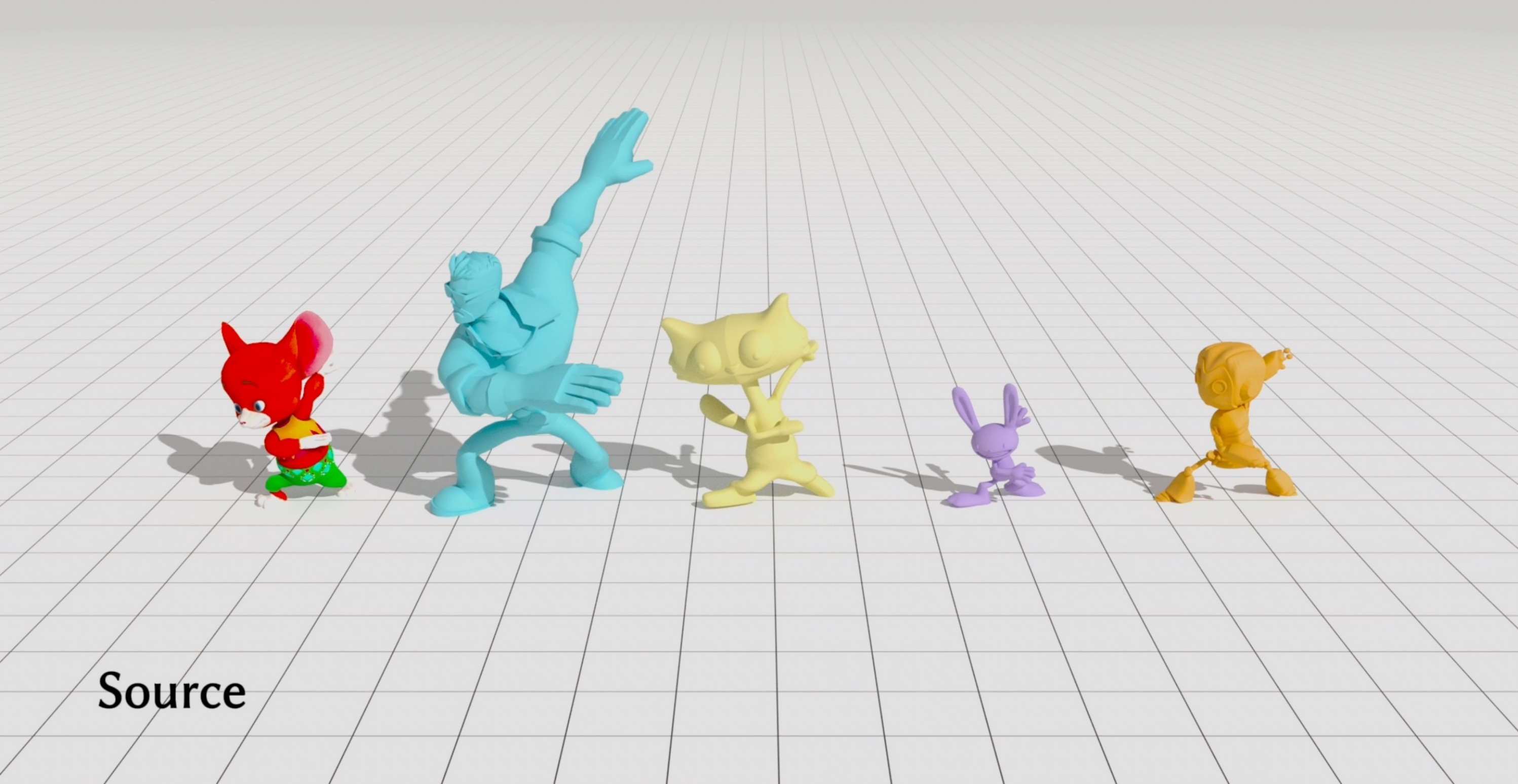}
\caption{\textbf{Qualitative retargeting to external OOD characters.} Our method preserves the source action across large variations in scale, proportions, and overall morphology.}
\label{fig:ood}
\end{figure}
}

\modified{
\begin{figure}
\centering
\includegraphics[width=0.46\textwidth]{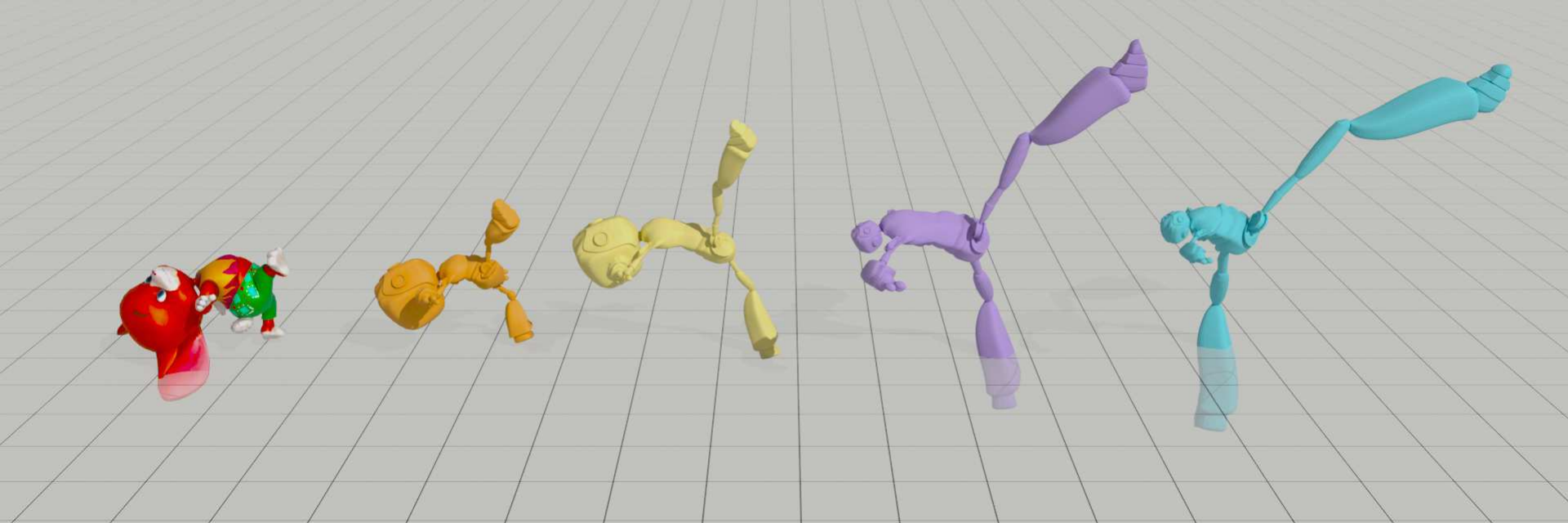}
\caption{\textbf{Retargeting morphology stress test.} The leftmost is the original Mixamo motion. From orange to blue, the target character is stretched from none to extreme. The ground plane is made semi-transparent, so the penetrations are seen.}
\label{fig:ood_stress_test}
\end{figure}
}

%% file: conclusion.tex
\section{CONCLUSION AND LIMITATIONS}

We presented a unified conditional-flow framework for text-to-motion generation, zero-shot text-based editing, and zero-shot intra-structural retargeting. A single rectified-flow model conditioned on text and skeleton supports all three tasks through a shared inference rule and residual multi-condition classifier-free guidance, reducing the need for separate generation, editing, and skeleton-adaptation pipelines.

Our current scope remains skeleton-level, intra-structural retargeting. Within this scope, the full pipeline achieves the best MSE and Contact Accuracy in our evaluation, with substantially lower foot-skate speed than prior baselines. It still does not address cross-topology transfer, mesh-aware retargeting, or explicit physical control, and difficult cases can show floor penetration or body-part interpenetration under large morphology changes. See representative failure cases in the supplementary. Our guidance analysis further indicates that retargeting remains skeleton-dominated, with text contributing only a weak residual effect. Strengthening semantic--skeletal coupling and incorporating geometry-aware objectives remain important directions beyond the present scope.

%% file: images.tex
\begin{figure*}
    \centering
    \includegraphics[width=\linewidth]{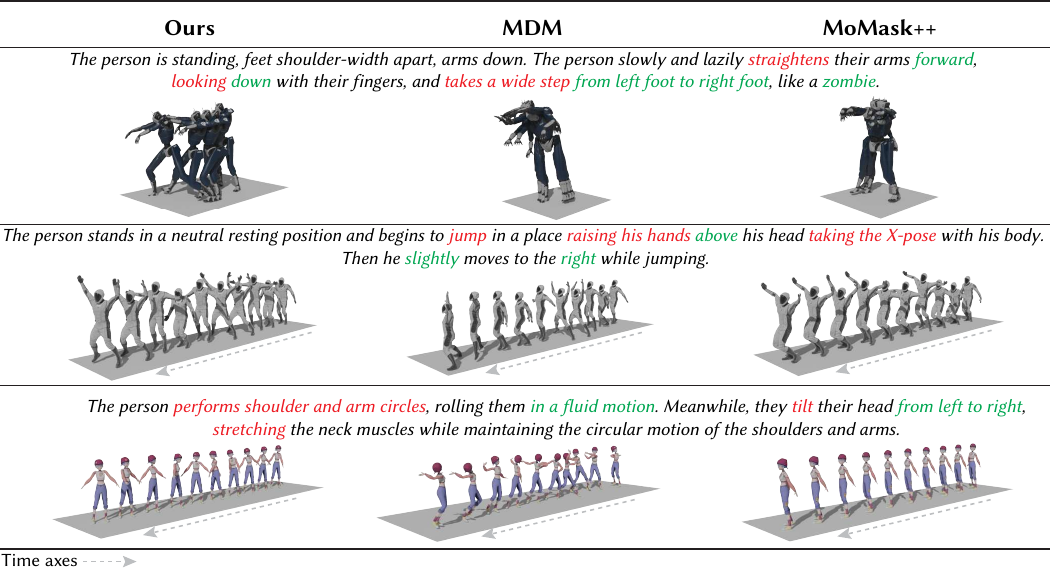}
    \caption{Qualitative comparison on text-to-motion generation. For visualization, motions with little or no root translation are manually time-shifted. 
    Prompt words in red denote actions, while words in green indicate motion modifiers.
    Our model successfully synthesizes coherent full-body motions that faithfully reflect fine-grained textual details over long time spans. This temporal coherence is critical for our downstream editing tasks, where the model must preserve the narrative structure of the source motion.}
    \label{fig:result-gen}
\end{figure*}

\begin{figure*}
\centering
\includegraphics[width=\linewidth]{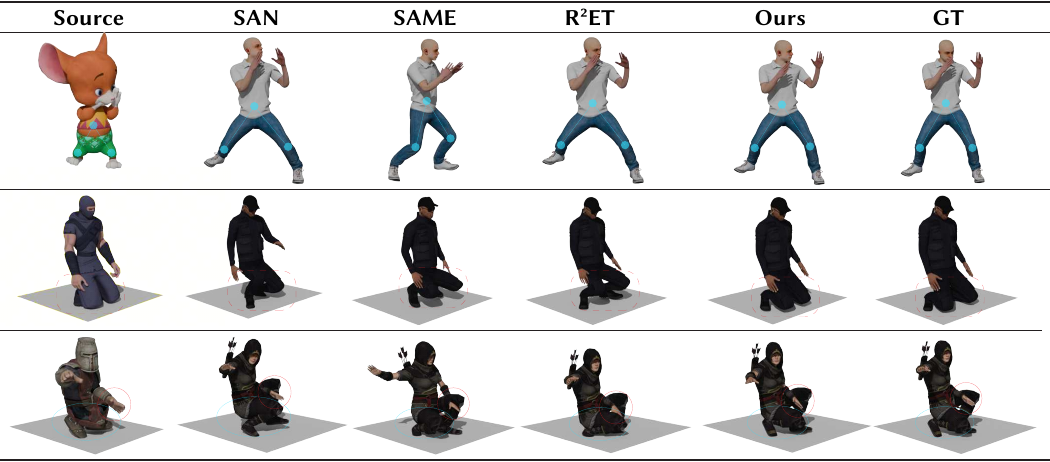}
\caption{Qualitative retargeting comparison. We compare against SAN \cite{aberman2020san}, SAME \cite{lee2023same}, and R2ET \cite{zhang2023r2et}. 
The proposed method better preserves fine-grained local motion and adapts to varying skeleton proportions.
As highlighted in the zoomed-in regions, baseline methods often introduce artifacts such as unnatural twisting or over-stretching when adapting to different skeletons. In contrast, our method preserves delicate local details (e.g., leg bending scale, hand facing direction) while naturally adapting the global motion to fit the new body shape.}
\label{fig:result-retarget}
\end{figure*}

\begin{figure*}
    \centering
    \includegraphics[width=\linewidth]{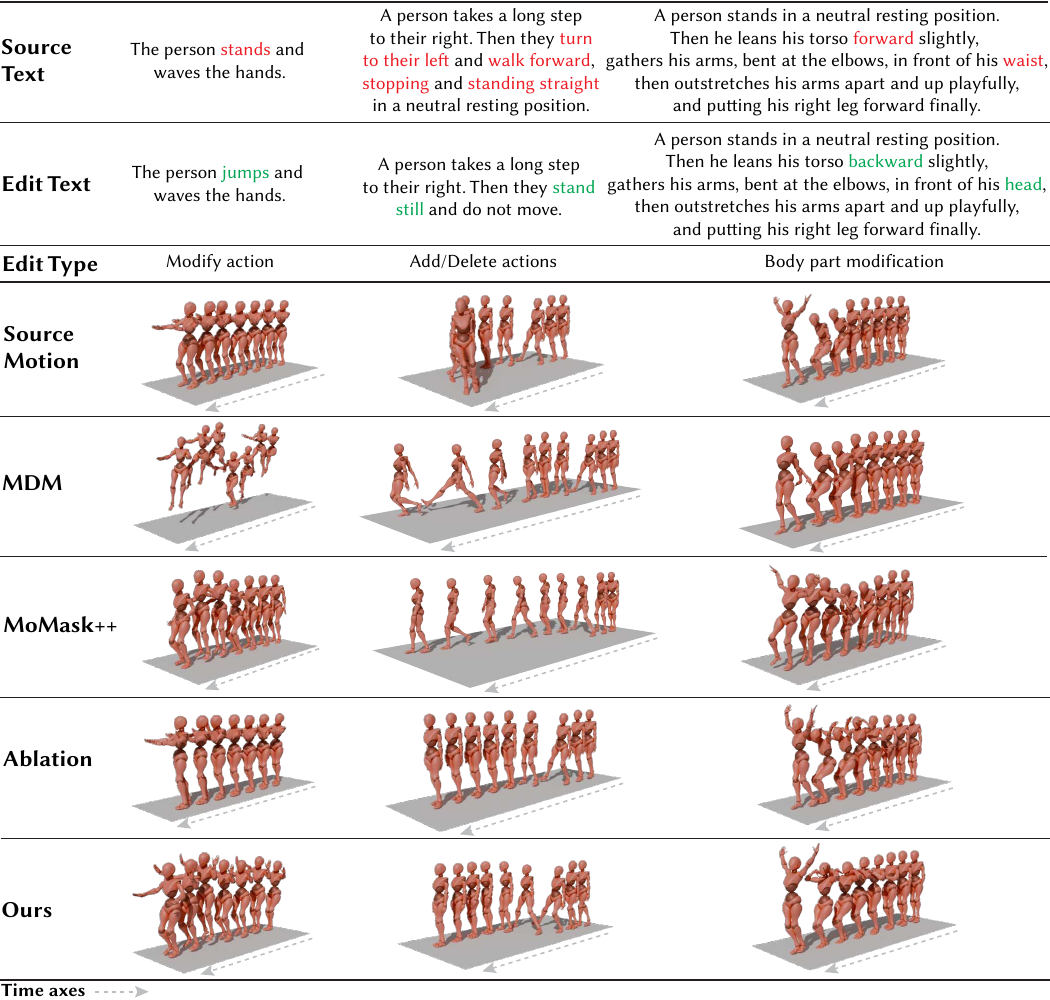}
    \caption{Qualitative results on text-based motion editing. 
    Edited prompt words are highlighted in red and green. 
    For visualization, motions shifted as in Fig.~\ref{fig:result-gen}. 
    Our method demonstrates superior adaptability across all edit types. 
    Unlike mask-based approaches (MDM) which can introduce stiffness at mask boundaries, or regenerative baselines (MoMask++) which often drift from the source semantics, Our method produces accurate, coherent edits.}
    \label{fig:result-edit}
\end{figure*}